# A transferable framework for structure–energy mapping of nanovoid–solute complexes: Tungsten alloys as a model system

Kang-Ni He[a,b], Xiang-Shan Kong[a,b*], Jie Hou[c], Chang-Song Liu[d], Zhuo-Ming Xie[d*]

[a] State Key Laboratory of Advanced Equipment and Technology for Metal Forming, Shandong University, Jinan 250061, China

[b] Key Laboratory for Liquid-Solid Structural Evolution and Processing of Materials (Ministry of Education), Shandong University, Jinan 250061, China

[c] College of Materials Science and Engineering, State Key Laboratory of Cemented Carbide, Hunan University, Changsha 410082, China

[d] Key Laboratory of Materials Physics, Institute of Solid State Physics, HFIPS, Chinese Academy of Sciences, Hefei 230031, China

**Abstract**

Understanding the structures and energetics of nanovoid–solute complexes is essential for elucidating the coupled evolution of defects in metals. Yet their vast and complex configurational space poses a major challenge to conventional approaches. Using W–Re as a representative system, we demonstrate that solute segregation at nanovoid surfaces can be decomposed into direct nanovoid–solute interactions and nanovoid-mediated solute–solute interactions. Both are governed by local coordination motifs, with identical motifs giving nearly identical energetics. Based on first-principles data, we trained machine-learning models to map diverse local motifs to their energetics, enabling the energetics of any nanovoid–solute complex to be reconstructed from a finite set of constituent local motifs. We further developed a size-dependent configurational-search framework to efficiently identify thermodynamically stable structures, using exhaustive enumeration, simulated annealing, and greedy addition for small, medium-sized, and large complexes, respectively. This framework enabled the construction of a large database, revealed the staircase-like segregation behavior of Re, and derived a simple criterion based on Re surface coverage for rapid energy prediction across a wide size range. It also links Re segregation to vacancy-mediated nanovoid evolution and provides benchmarks for existing models and empirical potentials. Extensions to Os and Ta support the generality of the local-motif concept, and the predicted segregation behavior of solutes at nanovoids agrees with a range of experimental observations. This work establishes a physically transparent, accurate, and transferable framework for studying nanovoid–solute co-evolution in metals and provides reliable energetic inputs for multiscale simulations.

**Keywords:** Nanovoid–solute complexes; Tungsten–rhenium alloys; Local coordination motifs; Configurational search; Defect evolution

# 1. Introduction

Vacancies are the most fundamental point defects in metals and can be generated in large excess under non-equilibrium conditions such as irradiation, rapid quenching, or plastic deformation. At sufficiently high concentrations, vacancies aggregate and progressively evolve into nanovoids. Solute atoms, whether intentionally introduced to tailor material properties or inadvertently introduced during processing or service, often interact strongly with vacancies [1-9]. Generally, through vacancy-drag diffusion, solute atoms are continuously transported toward developing nanovoids, where they preferentially segregate at nanovoid surfaces, forming solute-enriched layers and, in some cases, inducing second-phase precipitation or core-shell structures [5-7, 10-14]. In turn, solute decoration modifies the thermodynamic stability and kinetic behavior of vacancies and nanovoids, thereby regulating nanovoid nucleation, growth, and evolution [12-14]. Such coupled nanovoid–solute phenomena have been widely observed in metallic systems including Fe-based [1-3, 15-17], W-based [4-6, 10-13], Al-based [7-9, 18, 19], Ni-based [14, 20, 21], and Cu-based [22-24] materials, and play a critical role in microstructural evolution and property changes under irradiation, thermal exposure, and mechanical loading, manifesting as hardening, embrittlement, void swelling, and reduced thermal conductivity.

Despite extensive experimental observations of solute-decorated nanovoids, which have provided valuable insights into their overall morphology and compositional distributions [5-7, 10-12], direct access to the real-time atomic-scale structures and energetic characteristics of nanovoid–solute complexes during their formation and evolution remains severely limited. This limitation arises primarily from the intrinsic constraints in temporal and spatial resolution of current experimental techniques, particularly under extreme conditions such as irradiation or high-temperature exposure. Consequently, atomistic simulations, including first-principles calculations and molecular dynamics/statics (MD/MS) simulations, have become indispensable tools for probing the fundamental physics of nanovoid–solute complexes [25-39]. In a typical computational framework, multiple initial configurations consisting of vacancies and solute atoms are artificially constructed and subsequently relaxed using first-principles calculations or MD/MS simulations. Stable configurations are then identified, and key energetic descriptors, such as formation energies, binding energies, and migration barriers, are evaluated. This strategy has generated a substantial body of quantitative atomistic data for small nanovoid–solute complexes, usually comprising only a few vacancies and solute atoms, thereby significantly advancing the understanding of cluster nucleation and early-stage evolution.

However, in realistic materials, such small complexes rarely remain isolated and stable. Instead, they dynamically evolve through growth and dissociation, for example via Ostwald ripening [40, 41], wherein smaller complexes dissolve while larger ones grow. This process leads to the emergence of medium- and large-sized nanovoid–solute complexes with characteristic dimensions of several nanometres, involving hundreds to thousands of solute atoms and vacancies [5-7, 10-12]. With increasing complex size, the number of possible atomic configurations undergoes a combinatorial explosion, rendering the traditional "enumerate-and-relax" paradigm computationally prohibitive in terms of both computational cost and data management. Consequently, systematic and quantitative structure–energy databases for medium- and large-sized nanovoid–solute complexes remain largely unavailable. This critical data gap severely limits the development of physically grounded models capable of quantitatively predicting solute segregation, nanovoid growth kinetics, and their coupled evolution across length and time scales.

To address this challenge, a series of modeling strategies have been developed, among which cluster-expansion (CE)-based and related lattice-Hamiltonian models [42, 43] represent a widely adopted framework for systematically studying vacancy–solute interactions in W, particularly for transmutation elements such as Re and Os [44-46]. The relevance of this system stems from the fact that tungsten is a leading candidate material for plasma-facing components in fusion reactors, where it is subjected to intense neutron irradiation. Under such extreme conditions, large concentrations of vacancies are continuously generated, while transmutation reactions produce significant amounts of Re and Os. The subsequent coupled evolution of vacancies and transmutation solutes leads to the formation of vacancy clusters and nanovoids decorated by Re/Os, which can substantially alter the mechanical and thermophysical properties of tungsten. Within this context, Wróbel et al. [44] constructed an Ising-type cluster expansion Hamiltonian based on extensive density functional theory (DFT) datasets, and combined it with Monte Carlo (MC) simulations to predict finite-temperature solute segregation behavior and the thermodynamic stability of Re–vacancy and Re–nanovoid complexes. Building on this foundation, Nguyen-Manh et al. [45] extended the CE formalism to multicomponent W–(Re, Os, Ta)–vacancy systems by incorporating DFT-derived cluster interactions and employing thermodynamic integration techniques, enabling a more comprehensive description of temperature-dependent free energies and phase stability in chemically complex environments.

Further progress was made by Lloyd et al. [46], who introduced an extended Ising-like Hamiltonian in which interaction parameters explicitly depend on local solute and vacancy concentrations. By parameterizing the model using DFT-calculated mixing enthalpies, this approach captures nonlinear solute–vacancy coupling effects more accurately. When implemented in Monte Carlo simulations, it significantly improves the predictive capability for composition evolution and vacancy accumulation under irradiation conditions. In addition, Huang et al. [47, 48] developed a CE-based lattice Hamiltonian for the W–Re system and systematically evaluated the binding energetics of small vacancy–Re clusters. Importantly, they derived two analytical expressions describing (i) the incremental binding energy upon adding a Re atom to an existing vacancy–Re cluster, and (ii) the incremental binding energy upon adding a vacancy to the same cluster. These expressions have subsequently been widely adopted in object kinetic Monte Carlo (OKMC) simulations [49, 50], enabling large-scale modeling of irradiation-driven defect thermodynamics and kinetics in tungsten.

Despite these advances, CE-based approaches share a common conceptual structure in which configurational energetics are represented through a lattice Hamiltonian constructed from a finite set of effective cluster interactions. This formulation is well suited for bulk alloys and small defect clusters, where local environments can be adequately described within a compact cluster basis defined on the parent lattice. For nanovoid–solute systems, however, the coupled aggregation of vacancies and segregation of solutes gives rise to a broad spectrum of non-equivalent local structural environments, even though the underlying crystalline lattice remains preserved. In such cases, identical cluster motifs may correspond to physically distinct configurations depending on their embedding atomic surroundings. As a result, the mapping between cluster functions and local energetic contributions becomes increasingly non-unique, making it difficult for a set of cluster interactions to provide a consistent and transferable representation of configurational energetics across different nanovoid sizes and compositions. While, in principle, extending the expansion to higher-order clusters and longer interaction ranges can improve the representation, this leads to a combinatorial growth in the number of configurations and associated training data, posing significant practical challenges for defect-rich systems.

These considerations suggest that, for nanovoid–solute complexes, it is advantageous to adopt a formulation in which configurational energetics are described more directly in terms of physically identifiable local structural environments, rather than being implicitly encoded in a lattice-based cluster expansion. This perspective motivates the development of an alternative framework in which the energetics are constructed from elementary local processes associated with solute adsorption at nanovoid surfaces. In this work, using W–Re as a representative system, we develop a physically transparent and scalable framework to investigate the structures and energetics of nanovoid–solute complexes. We show that solute segregation at nanovoid surfaces can be rigorously decomposed into direct nanovoid–solute interactions and nanovoid-mediated solute–solute interactions, both of which are governed primarily by local coordination motifs. Based on first-principles calculations, we establish machine-learning models that map local motifs to their corresponding energetics, thereby enabling efficient reconstruction of the energetics of arbitrary nanovoid–solute complexes. To identify thermodynamically stable structures across different size regimes, we further develop a size-dependent configurational-search strategy based on three algorithms: exhaustive enumeration, simulated annealing, and greedy addition. On this basis, we construct a large structure–energy database, reveal the staircase-like segregation behavior of Re, derive a simple coverage-based criterion for rapid energy prediction, and assess the performance of existing models and empirical potentials. Extensions to Os and Ta further demonstrate the generality of the present framework.

## 2. Computational methods

### 2.1. First-principles DFT calculations

DFT calculations were performed using the Vienna Ab initio Simulation Package (VASP) [51-53]. The electron–ion interactions were described using the projector-augmented wave (PAW) method [54], and the exchange–correlation functional was treated within the generalized gradient approximation (GGA) using the Perdew–Burke–Ernzerhof (PBE) parameterization [55]. To balance computational cost and accuracy, two sets of body-centered cubic (bcc) supercells containing 250 and 432 atoms were employed, corresponding to $5 \times 5 \times 5$ and $6 \times 6 \times 6$ replications of the conventional unit cell, respectively. The smaller supercells were used for vacancy–solute complexes composed of $n$ vacancies and $m$ solute atoms when $n \leq 6$, whereas the larger supercells were adopted for cases with $n > 6$. A plane-wave cutoff energy of 500 eV was adopted, which was sufficient to ensure convergence of the total energies. Brillouin-zone integrations were performed using a $3 \times 3 \times 3$ Monkhorst–Pack k-point mesh [56]. Partial occupancies were treated using the Methfessel–Paxton smearing scheme with a width of 0.2 eV. The atomic configurations, as well as the cell shape and volume, were fully relaxed until the total energy and atomic forces converged to $1 \times 10^{-5}$ eV and 0.01 eV/Å, respectively. All parameters were carefully tested to ensure convergence of the total energies. Structural visualizations were generated using VESTA [57].

2.2. Gradient boosting decision tree model

A gradient boosting decision tree (GBDT) model [58] was employed for training and prediction, implemented using the scikit-learn package [59]. To identify the optimal hyperparameters of the model, a grid-search strategy was adopted. The performance of each hyperparameter combination was evaluated using ten-fold cross-validation (10-fold CV) and the root-mean-square error (RMSE). The ranges of hyperparameters explored were as follows: learning rate $learning\_rate \in [0.05, 0.50]$ with a step size of 0.05; number of estimators $n\_estimators \in [50, 300]$ with a step size of 25; and maximum tree depth $max\_depth \in \{3, 5, 7\}$. All other parameters were kept at their default values in the software package. The training data were obtained from DFT calculations and compiled into the Supplementary data file.

2.3. Molecular statics simulations

Molecular statics (MS) simulations were carried out using the Large-scale Atomic/Molecular Massively Parallel Simulator (LAMMPS) [60] to assess various empirical interatomic potentials [61-64]. A periodic cubic supercell comprising $16 \times 16 \times 16$ repetitions of the bcc conventional unit cell (totaling 8192 atoms) was employed to minimize spurious interactions between periodic images. For each potential, atomic positions and the simulation cell (both shape and volume) were relaxed under zero-pressure conditions.

**3. Results and discussion**

3.1. Structure–energy modeling of nanovoid–solute complexes

The thermodynamic stability of a substitutional solute adsorbed at the surface of a pre-existing vacancy–solute complex in a metallic matrix is governed primarily by two interactions. As illustrated in Fig. 1, the first is the interaction between the solute and the surrounding metal atoms forming the nanovoid framework, which controls the energetic stability of adsorption at the surface. The second is the interaction between the newly adsorbed solute and solutes already present on the nanovoid surface, which modifies the local chemical environment and thereby modulates the binding affinity and spatial distribution of solutes. Together, these interactions determine the preferred adsorption sites near the nanovoid.

To quantify this process, we define the incremental binding energy for adsorption of the $m$-th solute atom onto a $V_n S_{m-1}$ complex, where $V_n S_{m-1}$ denotes a vacancy–solute complex containing $n$ vacancies and $m-1$ solute atoms, as

$$E_b(V_n S_{m-1}, S_1) = [E_t(S_1) + E_t(V_n S_{m-1})] - [E_t(V_n S_m) + E_t(\text{perf})], \quad (1)$$

where $V$ and $S$ represent vacancies and solute atoms respectively, with subscripts $n$ and $m$ denoting their quantities. The $E_t$ terms represent the total energies of the supercells containing a single substitutional solute atom, the $V_n S_{m-1}$ complex, the $V_n S_m$ complex, and the perfect crystal, respectively. A positive binding energy indicates that adsorption of the solute atom is thermodynamically favorable, whereas a negative value indicates an unfavorable interaction.

This incremental binding energy can be decomposed into two contributions

$$E_b(V_n S_{m-1}, S_1) = E_b(V_n, S_1) + E_b(S_{m-1}, S_1), \quad (2)$$

where

$$E_b(V_n, S_1) = [E_t(S_1) + E_t(V_n)] - [E_t(V_n S_1) + E_t(\text{perf})], \quad (3)$$

captures the interaction between the solute atom and the pristine nanovoid, and

$$E_b(S_{m-1}, S_1) = [E_t(V_n S_1) + E_t(V_n S_{m-1})] - [E_t(V_n S_m) + E_t(V_n)], \quad (4)$$

represents the interaction between the newly adsorbed solute and those already bound to the nanovoid.

Summing these contributions over all solute atoms yields the total binding energy of the $V_n S_m$ complex

$$E_B(V_n, S_m) = \sum_{i=1}^{m} E_b(V_n, S_1^i) + \frac{1}{2} \sum_{i=1}^{m} E_b(S_{m-1}, S_1^i), \quad (5)$$

where $S_1^i$ denotes the $i$-th solute atom. The first term represents the solute–nanovoid interactions, whereas the second term represents the solute–solute interactions. The factor of 1/2 is introduced to avoid double counting. For a given composition, a larger total binding energy indicates stronger overall stabilization and hence greater thermodynamic stability of the $V_n S_m$ complex.

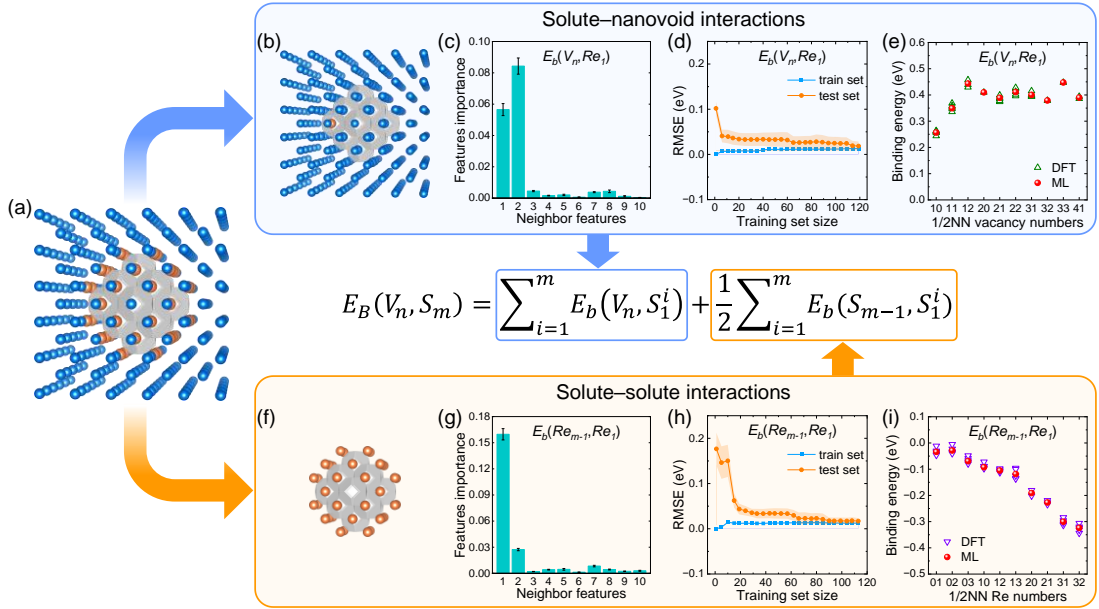

**Fig. 1.** Structure–energy modeling of nanovoid–solute complexes. (a) Schematic of a nanovoid–solute complex in a metallic matrix. The gray truncated octahedra denote vacancies, whereas the blue and orange spheres represent matrix and solute atoms, respectively. (b) Schematic of the interaction between a nanovoid and a single solute atom. (c) Feature-importance analysis of different nearest-neighbor vacancy coordination features for $E_b(V_n, Re_1)$. (d) Learning curve for $E_b(V_n, Re_1)$ using the first- and second-nearest-neighbor (1–2NN) vacancy coordination features, with the shaded region indicating the standard deviation from cross-validation, as in the panels below. (e) Values of $E_b(V_n, Re_1)$ for different 1–2NN vacancy coordination numbers. (f) Schematic of the interactions among multiple solute atoms adsorbed on a nanovoid surface, with the matrix atoms omitted for clarity. (g) Feature-importance analysis of different nearest-neighbor Re coordination features for $E_b(Re_{m-1}, Re_1)$. (h) Learning curve for $E_b(Re_{m-1}, Re_1)$ using the 1–2NN Re coordination features. (i) Values of $E_b(Re_{m-1}, Re_1)$ for different 1–2NN Re coordination numbers.

Given the vast configurational space of vacancy clusters, this study focuses on the most stable structures identified by minimizing their Wigner–Seitz (WS) surface area [65], namely nanovoids formed by the compact aggregation of vacancies. Based on these optimized structures, two types of configurations were constructed to examine solute behavior near nanovoids: nanovoids containing a single solute atom to characterize nanovoid–solute interactions (see Fig. 1(b)), and nanovoids with multiple solutes adsorbed on the inner surface to probe nanovoid–mediated solute–solute interactions (see Fig. 1(f)). Owing to the strong locality of atomic interactions, solute stability is assumed to depend primarily on the local coordination environment, described by the numbers of neighboring vacancies and solutes within a cutoff corresponding to the first ten neighbor shells. DFT calculations of vacancy–Re and Re–Re binding energies (see Supplementary Fig. S1) support this locality assumption and justify the cutoff choice. Using this coordination-based classification, 137 inequivalent configurations were identified for single-solute nanovoids (≤25 vacancies) and 131 for multiple-solute nanovoids (≤30 total defects). All configurations were fully relaxed by DFT calculations, generating a dataset of incremental and total binding energies together with the corresponding local coordination descriptors.

Table 1. Optimal hyperparameters and 10-fold cross-validated performance of gradient boosting decision tree models. Hyperparameter values in each cell are listed left to right as learning rate, number of estimators and maximum tree depth. All other parameters remain at the default values of scikit-learn. RMSE is reported in electronvolts (eV).

| Target | 1–10NN features | | 1–2NN features | |
|---|---|---|---|---|
| | hyperparameters | RMSE | hyperparameters | RMSE |
| $E_b(V_n, Re_1)$ | 0.05; 75; 5 | 0.019 | 0.50; 175; 3 | 0.020 |
| $E_b(Re_{m-1}, Re_1)$ | 0.35; 50; 3 | 0.018 | 0.50; 300; 3 | 0.019 |

Notably, the configurational space of possible coordination environments cannot be exhaustively sampled. Therefore, machine-learning models were constructed to predict the energetics of unseen configurations. Based on this structure–energy dataset, two GBDT models were trained to predict nanovoid–Re binding energies, $E_b(V_n, Re_1)$, from vacancy-coordination descriptors and Re–Re interaction energies, $E_b(Re_{m-1}, Re_1)$, from solute-coordination descriptors (hyperparameters in Table 1). Feature-importance analysis shows that the dominant contributions arise from the first- and second-nearest-neighbor coordination features (1–2NN), with negligible influence from more distant neighbors (see Fig. 1(c) and (g)). Learning-curve results show that models trained using only the 1–2NN features perform comparably to those using the 1–10NN feature set, while exhibiting no evidence of overfitting (see Fig. 1(d) and (h), as well as Supplementary Fig. S2), confirming that the first two neighbor shells sufficiently capture the relevant local environment. More importantly, Re atoms occupying equivalent 1–2NN environments exhibit nearly identical binding energies, independent of nanovoid size or total Re content (see Fig. 1(e) and (i); the corresponding configurations are shown in Supplementary Fig. S3). Within the present dataset, only 14 and 24 distinct local environments are found for the above two structure types, respectively (see Supplementary Table S1 for all types and the corresponding energetic data), and they account for all coordination types in the most stable nanovoids containing up to 300 vacancies. For larger or metastable nanovoids, new coordination types arise, and the trained GBDT models provide a quantitative surrogate capable of rapidly predicting their incremental binding energetics without additional DFT calculations.

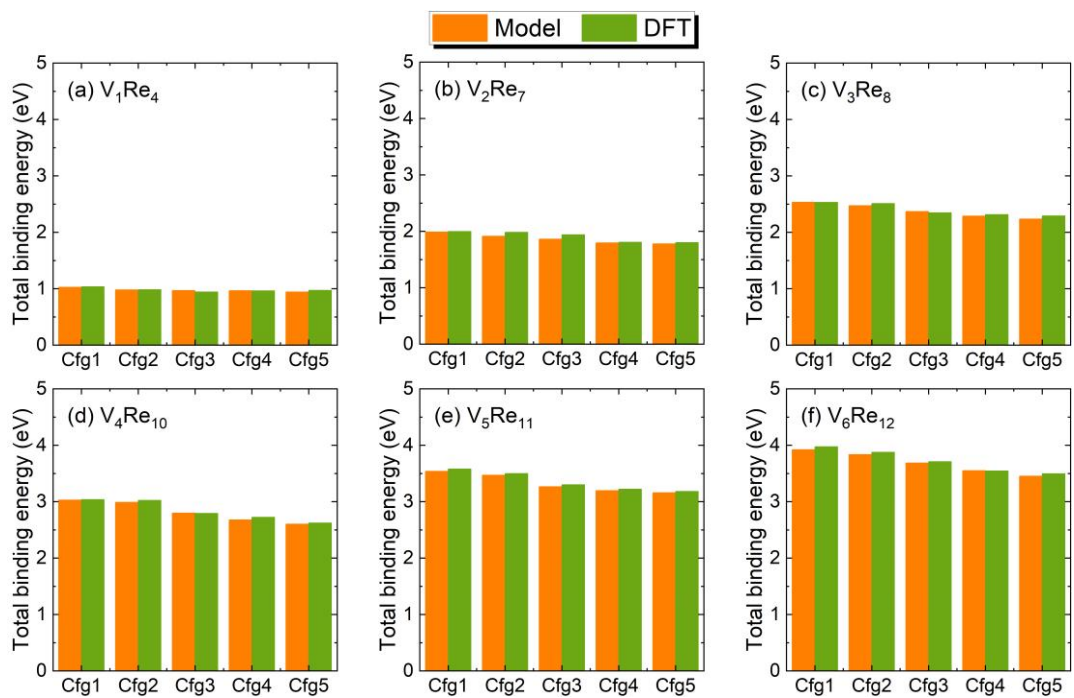

**Fig. 2.** Comparison between model-predicted and DFT-calculated total binding energies for selected configurations of nanovoid–Re complexes. (a–f) Results for $V_1Re_4$, $V_2Re_7$, $V_3Re_8$, $V_4Re_{10}$, $V_5Re_{11}$, and $V_6Re_{12}$, respectively. For each complex, the most stable configuration, corresponding to the largest total binding energy, and four metastable configurations were selected from the exhaustively enumerated structures; the specific configurations are shown in Supplementary Fig. S4.

Accordingly, any nanovoid–Re complex can be represented as an assembly of a finite set of distinct local-environment motifs. By mapping the local environment of each Re atom to the corresponding precomputed incremental binding energies, $E_b(V_n, Re_1)$ and $E_b(Re_{m-1}, Re_1)$, and summing these contributions, the total binding energy of a complex can be evaluated efficiently. This decomposition reduces the energy evaluation of large complexes to a summation of independent and reusable local contributions, thereby greatly simplifying the problem. To validate the approach, six nanovoid–Re complexes with varying numbers of vacancies and Re atoms (i.e. $V_1Re_4$, $V_2Re_7$, $V_3Re_8$, $V_4Re_{10}$, $V_5Re_{11}$, and $V_6Re_{12}$) were examined. For each complex, all possible Re site occupations were exhaustively enumerated, and total binding energies were calculated using the local-environment model. For each $V_nRe_m$ complex, the most stable configuration and four metastable configurations were selected at prescribed sampling intervals, with the specific structures shown in Supplementary Fig. S4. These complexes were then fully relaxed using DFT calculations to obtain their total binding energies. As shown in Fig. 2, the model-predicted total binding energies are in excellent agreement with the DFT results, confirming the high accuracy of the approach and further validating the local-coordination-based framework.

3.2. Thermodynamic stability framework for nanovoid–solute complexes

Building on the predictive model for evaluating total binding energies of arbitrary nanovoid–solute complexes, the most stable configuration, corresponding to the maximum total binding energy, can be identified by searching the configurational space of possible solute occupations. For a vacancy cluster $V_n$ with $q$ potential solute sites, the number of possible configurations containing $m$ solute atoms is

$$N_{EE} = \binom{q}{m} = \frac{q!}{m!(q-m)!}. \tag{6}$$

This number grows combinatorially with cluster size, leading to a rapid expansion of the configurational space. To address this challenge, we develop a size-dependent configurational search framework that employs different optimization strategies for different system sizes in order to balance computational efficiency and accuracy.

For small complexes, exhaustive enumeration (EE) provides an exact method for identifying the global minimum. However, as cluster size increases, the combinatorial growth of $N_{EE}$ quickly renders EE impractical. In such cases, the simulated annealing (SA) algorithm provides a practical alternative. The workflow of SA is shown in Fig. 3(a). In each step, a trial move is generated by randomly picking one solute and one host atom on the $V_n$ surface and attempt a swap. Moves that increase the total binding energy are accepted, whereas unfavorable moves are accepted with a probability determined by the Metropolis criterion. This stochastic sampling enables efficient exploration of the configurational landscape and promotes convergence toward energetically favorable configurations. Its computational cost scales as

$$N_{SA} = n \log_\alpha \left(\frac{T_2}{T_1}\right), \tag{7}$$

where $T_1$ and $T_2$ are the initial and final temperatures, respectively, $\alpha$ is the cooling rate, and $n$ is the number of iterations at each temperature step. Compared with EE, the computational cost of SA scales much more favorably, providing an efficient means to explore the configurational landscape and converge toward energetically favorable structures in moderately sized complexes. We have previously successfully applied this algorithm to determine stable nanovoid structures [65] and the corresponding hydrogen adsorption configurations and energies [66], thereby confirming its robustness.

Although SA substantially reduces the search cost relative to EE and efficiently identifies energetically favorable configurations for moderately sized complexes, the combinatorial growth of possible configurations renders convergence toward the global optimum increasingly challenging for large nanovoid–solute complexes. To address this limitation, we introduce a greedy addition (GA) algorithm, as illustrated in the flowchart shown in Fig. 3(b), in which solute atoms are added sequentially to the site that maximizes the incremental binding energy. For a system with $q$ potential solute sites and $m$ solute atoms, the total number of configurations evaluated in GA is

$$N_{GA} = \sum_{k=0}^{m-1}(q-k) = mq - \frac{m(m-1)}{2}. \tag{8}$$

This approach substantially reduces the number of configurations that must be evaluated during the search for the most stable configuration, making GA particularly well suited for large nanovoid–solute complexes. A quantitative comparison of the computational costs of EE, SA, and GA is provided in the Supplementary Fig. S5. To assess the accuracy of GA, we compared the total binding energies of the most stable configurations predicted by GA with those obtained from EE and SA for a series of representative complexes (Fig. 3(d)). For these complexes, the results obtained from SA were nearly identical to those from EE, whereas GA yielded total binding energies slightly lower than those from EE, with an average deviation of approximately 2%. These results indicate that GA maintains high accuracy while greatly improving computational efficiency, thereby providing a reliable and scalable approach for high-throughput evaluation of large nanovoid–solute complexes in close agreement with more exhaustive methods. Within this framework, the combined use of EE, SA, and GA enables efficient exploration of the rapidly expanding configurational space and provides a scalable strategy for identifying thermodynamically stable nanovoid–solute structures across a wide range of complex sizes.

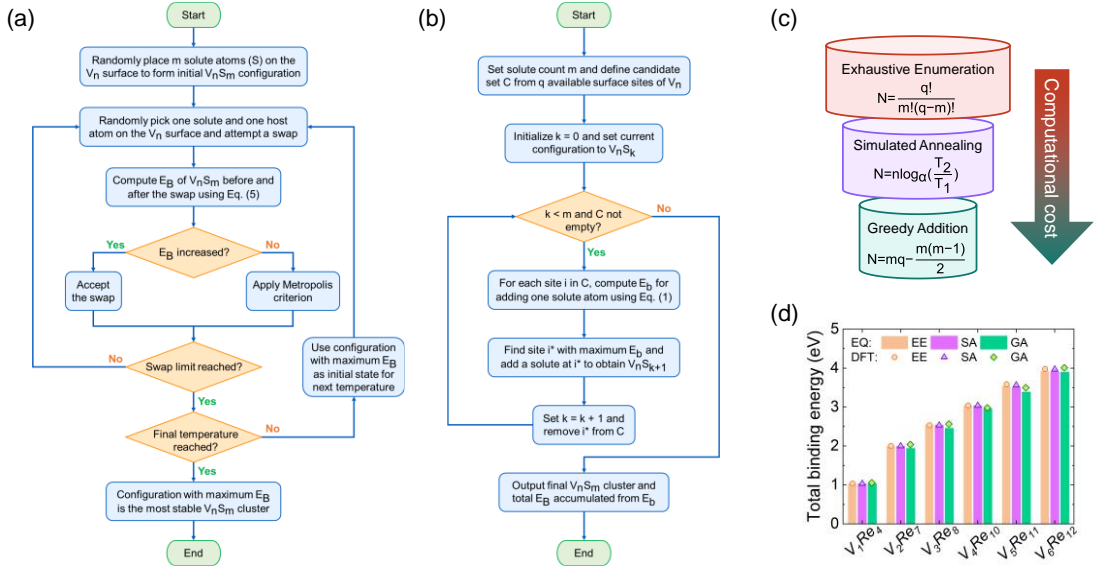

**Fig. 3.** Comparison of optimization algorithms for identifying stable nanovoid–solute complexes. (a) Flowchart of the simulated annealing (SA) algorithm. (b) Flowchart of the greedy addition (GA) algorithm. (c) Comparison of the computational costs of exhaustive enumeration (EE), SA, and GA. For a vacancy cluster $V_n$ with q potential solute sites, the number of possible configurations containing $m$ solute atoms is compared with the costs of SA and GA; $T_1$ and $T_2$ denote the initial and final temperatures, respectively, α is the cooling rate, and n is the number of iterations at each temperature step. (d) Comparison of the performance of EE, SA, and GA in identifying the $V_n Re_m$ configuration with the maximum total binding energy. Six nanovoid–Re complexes, $V_1 Re_4$, $V_2 Re_7$, $V_3 Re_8$, $V_4 Re_{10}$, $V_5 Re_{11}$, and $V_6 Re_{12}$, are examined, with DFT results included for reference.

3.3. Database construction and a rapid-prediction criterion

Using the above workflow, we systematically determined the stable structures and corresponding binding energies for nanovoid–Re complexes in which Re atoms occupy surface sites of the nanovoids. The resulting dataset spans nanovoids from $V_1$ to $V_{300}$, hereafter denoted as $V_{1-300} Re_m$, and comprises 46,794 distinct configurations. These results establish a comprehensive database of nanovoid–Re structures and energetics, which forms the basis for the subsequent thermodynamic analysis (see Supplementary data).

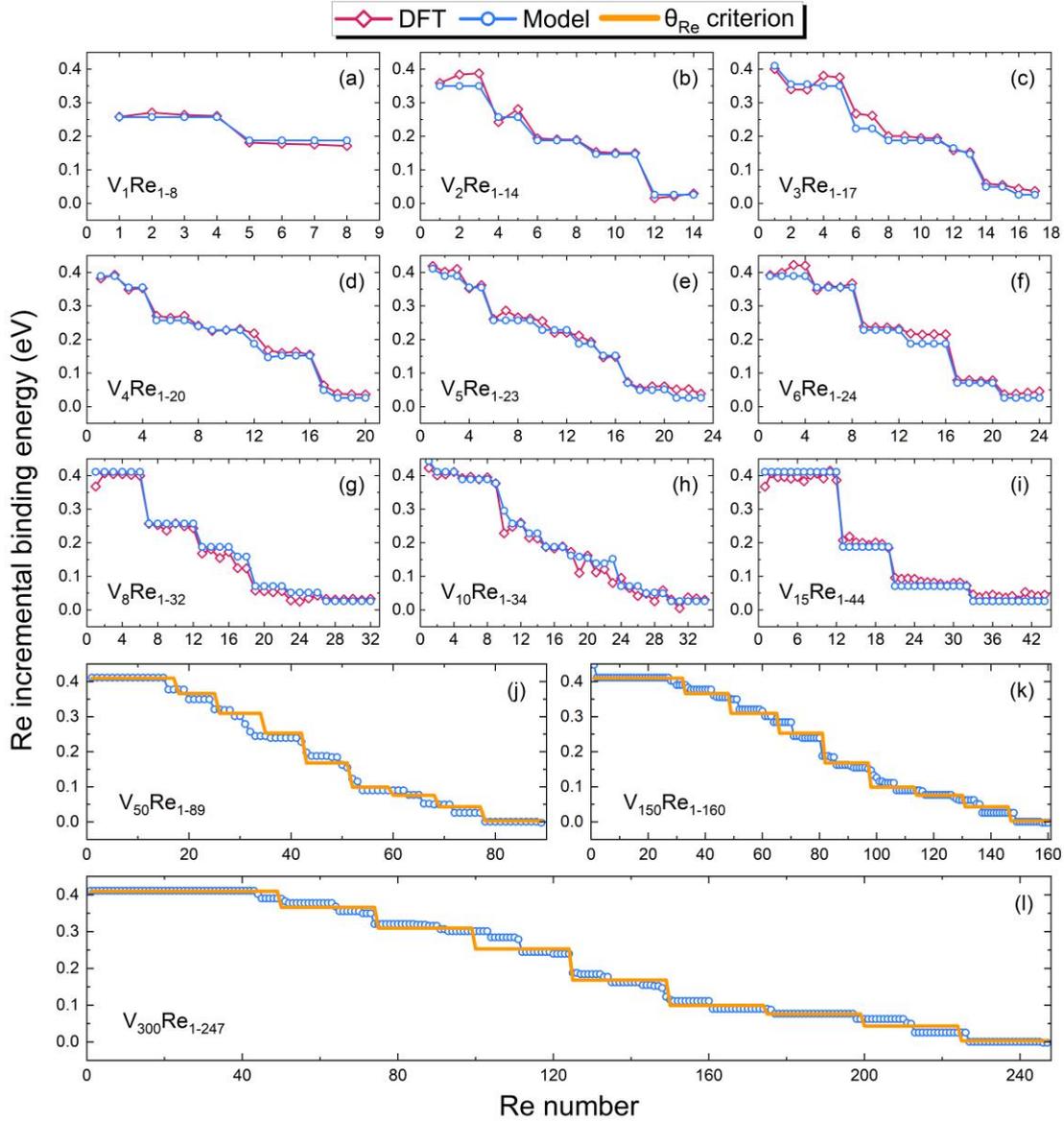

**Fig. 4.** Re incremental binding energies in nanovoid–Re complexes as a function of Re number. (a–i) Comparison between model-predicted and DFT-calculated values for $V_1 - V_6$, $V_8$, $V_{10}$, and $V_{15}$. (j–l) Comparison between the model-predicted values and those predicted by the $\theta_{Re}$-based criterion for larger nanovoids, $V_{50}$, $V_{150}$, and $V_{300}$.

Figures 4 and 5 illustrate the incremental binding energies of Re atoms and representative stable structures of nanovoid–Re complexes, respectively. In Fig. 4, the incremental binding energies predicted by the local-motifs model are directly compared with values obtained from fully relaxed DFT calculations for the same stable configurations. The two sets of results exhibit excellent agreement, thereby further validating the accuracy and predictive capability of the model. This consistency confirms that the essential energetics of Re adsorption on nanovoid surfaces are captured by the local coordination descriptors embedded in the framework, enabling reliable evaluation of large numbers of configurations without additional DFT calculations.

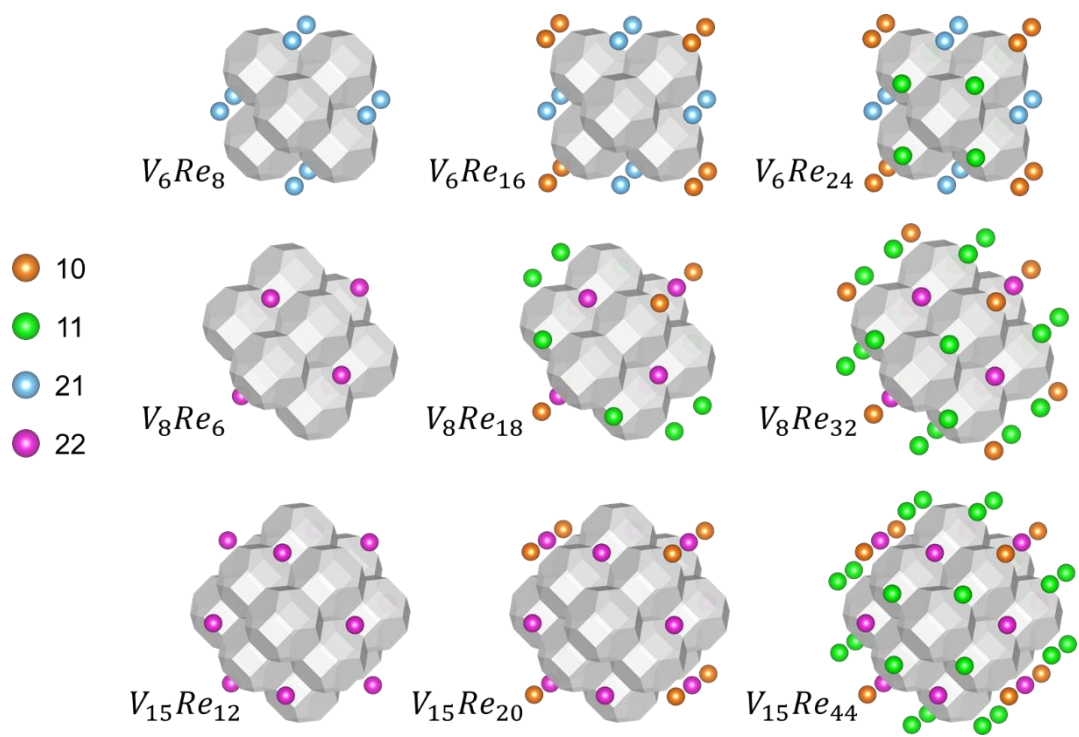

**Fig. 5.** Representative configurations of the highly symmetric nanovoid–Re complexes $V_6$, $V_8$, and $V_{15}$ at different Re contents. The gray truncated octahedron denotes the Wigner–Seitz polyhedron of a vacancy. Atoms in different colors correspond to surface sites with different numbers of first- and second-nearest-neighbor vacancies, as indicated in the legend on the left. For each two-digit label, the first and second digits denote the numbers of first- and second-nearest-neighbor vacancies, respectively; for example, "10" indicates 1 first-nearest-neighbor vacancy and 0 second-nearest-neighbor vacancies.

Notably, for nanovoids with a fixed vacancy number, the incremental binding energy of Re does not evolve smoothly with increasing Re content but instead exhibits a pronounced staircase-like behavior. According to Eq. (2), the incremental binding energy associated with adding one Re atom to a pre-existing $V_n Re_{m-1}$ complex can be decomposed into a nanovoid–Re interaction term and a nanovoid-mediated Re–Re interaction term. Because both contributions are governed by a limited number of local coordination environments, they assume only discrete values (see Fig. 1(e) and (i)), leading to a finite set of allowed incremental binding energies. This discrete energetic spectrum is directly reflected in the large-scale statistics of the database. Figure 6 summarizes the incremental binding energies of Re for all $V_{1-300}Re_m$ complexes as a function of Re number, and reveals a clear staircase-like decrease in energy. The result indicates that Re adsorption proceeds hierarchically, with surface sites being occupied in order of decreasing energetic favorability.

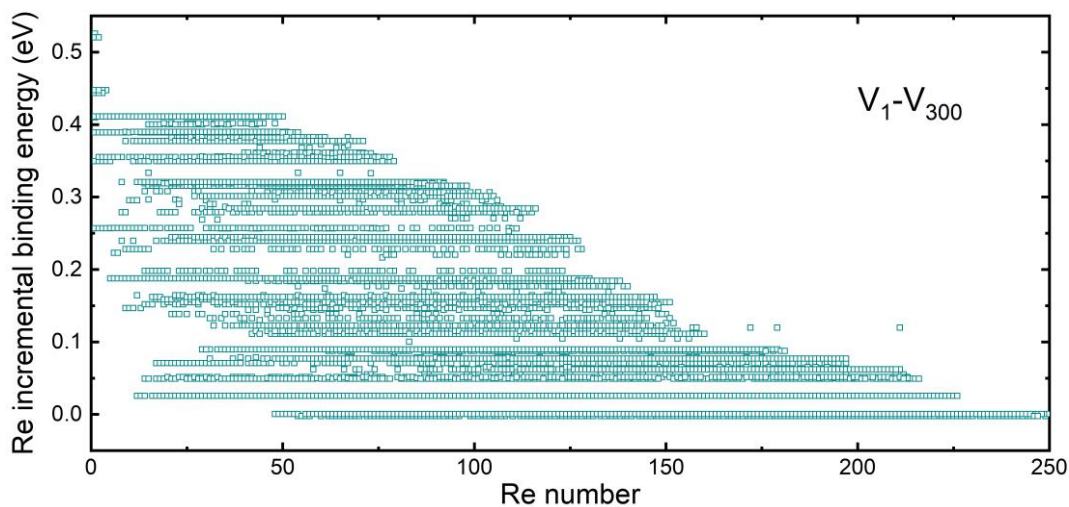

**Fig. 6.** Re incremental binding energies of all $V_{1-300}Re_m$ complexes as a function of Re number, predicted by the GA-based model.

At low Re coverage, Re atoms preferentially occupy energetically favorable surface sites characterized by large $E_b(V_n, Re_1)$ values. Simultaneously, repulsive Re–Re interactions favor spatially separated configurations, suppressing mutual solute interactions. Importantly, $E_b(V_n, Re_1)$ is highly concentrated and nearly uniform (0.40 ± 0.05 eV), with only minor deviations in exceptional cases, such as vacancy coordination type 10 (0.25 eV). Consequently, during the initial adsorption steps, the incremental binding energies are dominated by the nanovoid–Re interaction term and largely clustered around 0.40 eV (see Fig. 4). As coverage increases, however, the available separation between Re atoms becomes progressively constrained, and repulsive Re–Re interactions inevitably emerge. The corresponding $E_b(Re_{m-1}, Re_1)$ values span a much wider range (approximately 0 to -0.40 eV) than $E_b(V_n, Re_1)$, gradually dominating the incremental binding energy and driving the stepwise decrease observed in the staircase-like adsorption profile. The interplay between nearly uniform nanovoid–Re attraction at low coverage and strongly variable Re–Re repulsion at higher coverage governs the hierarchical adsorption sequence and shapes the overall energetic landscape of nanovoid–Re complexes.

Building on the hierarchical adsorption mechanism and the staircase-like evolution of incremental binding energies discussed above, the discrete nature of the local coordination environments implies that the incremental binding energy can assume only a finite set of allowed values. Importantly, the energy spacing between adjacent levels is small, such that many of these discrete states can be treated approximately as a single stage. Consequently, the evolution of the incremental binding energy with increasing Re coverage can be coarse-grained into a limited number of coverage-dependent stages, each represented by a characteristic energy. Guided by this coarse-grained perspective, we sought a concise formulation to capture the coverage-dependent evolution of Re incremental binding energies across nanovoids of different sizes. To this end, we introduced a normalized Re surface-coverage parameter

$$\theta_{Re} = \frac{m}{m_{max}}, \tag{9}$$

where $m$ is the number of Re atoms adsorbed on a given nanovoid and $m_{max}$ is the nanovoid's maximum Re capacity. The latter was determined by fitting the maximum-coverage data as a function of the vacancy number $n$ (Fig. 7(a)), yielding

$$m_{max}(n) = 5.26n^{2/3} + 14.24. \tag{10}$$

Within this framework, the incremental binding energy can be expressed as a piecewise function of $\theta_{Re}$. The problem then reduces to determining the appropriate number of sub-intervals, $N$, for representing the coverage dependence. To this end, the interval $\theta_{Re} \in (0,1]$ was uniformly divided into $N$ sub-intervals, with the mean incremental binding energy in each interval taken as a representative value. We systematically assessed how prediction accuracy depends on $N$. As shown in Fig. 7(b), increasing $N$ from small values rapidly improves both the root-mean-square error (RMSE) and the coefficient of determination ($R^2$). Beyond $N = 10$, however, further refinement yields only marginal gains, indicating that the essential coverage dependence has already been resolved. Balancing predictive accuracy against model simplicity, we therefore selected $N = 10$, corresponding to ten equal-width $\theta_{Re}$ intervals: (0,0.1], (0.1,0.2], …, (0.9,1.0]. All 46,794 incremental binding-energy data points were assigned to these intervals according to their $\theta_{Re}$ values, and the interval-averaged values were used as representative predictions. Notably, the mean values of the first two intervals differ by less than 0.01 eV and are effectively indistinguishable. These two intervals were thus merged, yielding a final nine-interval scheme (Fig. 7(c)). Under this coarse-grained representation, the overall RMSE is 0.02 eV and $R^2$=0.98, demonstrating that the essential energetic trends are retained with high fidelity. Figure 4(j–l) compares the incremental binding energies predicted by the model with those obtained using the $\theta_{Re}$-based criterion for representative nanovoids $V_{50}$, $V_{150}$, and $V_{300}$. In all cases, the $\theta_{Re}$-based criterion closely reproduces the model predictions. To further assess its extrapolative capability, we extended the comparison to substantially larger nanovoids, $V_{400}$, $V_{500}$, and $V_{600}$, whose vacancy numbers lie well beyond the range used to construct the descriptor (Fig. 7(d–f)). Remarkably, even in this regime, the $\theta_{Re}$-based predictions remain in close agreement with the model results.

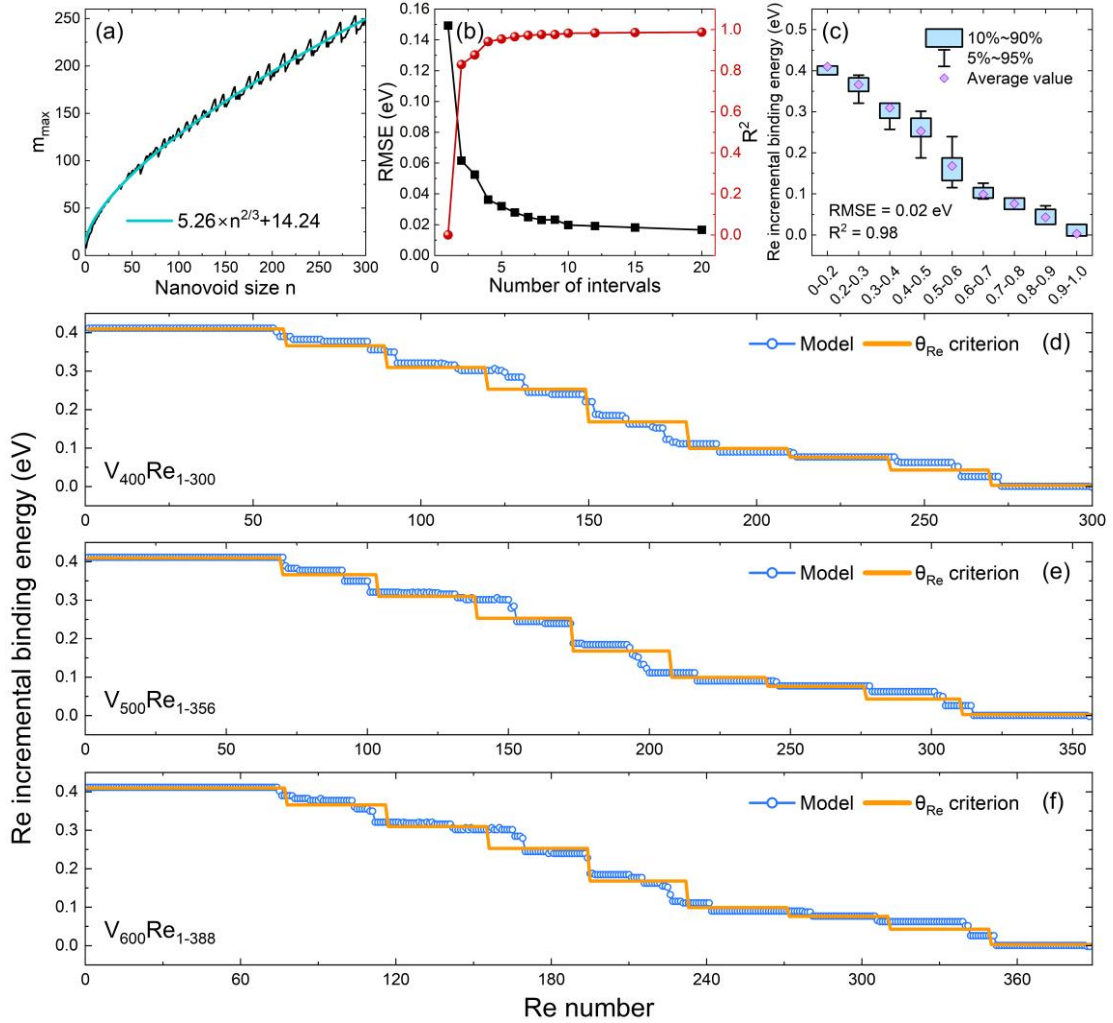

**Fig. 7.** Construction and validation of the $\theta_{Re}$-based criterion for predicting Re incremental binding energies, where $\theta_{Re} = m/m_{max}$ is the Re surface-coverage parameter. (a) Maximum Re capacity, $m_{max}$, as a function of nanovoid vacancy number n, together with the fitted relation. (b) Dependence of the prediction accuracy on the number of uniformly divided $\theta_{Re}$ intervals, measured by the root-mean-square error (RMSE) and coefficient of determination ($R^2$). (c) Final coarse-grained nine-interval scheme for Re incremental binding energies as a function of $\theta_{Re}$, based on all $V_{1-300}Re_m$ configuration and energy data. (d–f) Comparison between the predictive-model results and the $\theta_{Re}$-based predictions for larger nanovoids $V_{400}$, $V_{500}$, and $V_{600}$.

Overall, these results indicate that, although the configurational space of nanovoid–solute complexes cannot be exhaustively enumerated, the piecewise relation based on the normalized surface coverage $\theta_{Re}$, derived from the underlying local-coordination energetics, provides a physically grounded criterion for rapidly predicting the thermodynamic stability of nanovoid–Re complexes across arbitrary sizes.

3.4. Vacancy incremental binding energies of nanovoid–Re complexes

Building on the above framework, the stable structures and energetics of nanovoid–Re complexes with arbitrary nanovoid sizes can be predicted in a systematic and efficient manner. This capability enables a quantitative assessment of how Re segregation modifies the intrinsic thermodynamic stability and evolution of nanovoids. The growth and shrinkage of nanovoids are primarily governed by their ability to absorb and emit vacancies, which is conventionally characterized by the vacancy incremental binding energy. Specifically, the incremental binding energy associated with attaching a single vacancy to a $V_{n-1}Re_m$ complex to form $V_nRe_m$ is defined as

$$E_b(V_{n-1}Re_m, V_1) = [E_t(V_1) + E_t(V_{n-1}Re_m)] - [E_t(V_nRe_m) + E_t(\text{perf})], \quad (11)$$

where $E_t$ denotes the total energy of the corresponding configuration. After straightforward derivation, this quantity can be recast into a physically transparent form (see Supplementary Section S2 for the detailed derivation)

$$E_b(V_{n-1}Re_m, V_1) = E_b(V_{n-1}, V_1) + [E_B(V_n, Re_m) - E_B(V_{n-1}, Re_m)], \quad (12)$$

where $E_b(V_{n-1}, V_1)$ is the vacancy incremental binding energy for forming $V_n$ by attaching one additional vacancy to the pure nanovoid $V_{n-1}$, which can be obtained from our previously established WS-based predictive model. The second term explicitly quantifies the modification induced by Re segregation and is fully determined by the total nanovoid–Re binding energies $E_B(V_n, Re_m)$ and $E_B(V_{n-1}, Re_m)$ provided by the present model. This decomposition indicates that the influence of Re on nanovoid evolution enters primarily through its differential stabilization of successive nanovoid sizes. In essence, Re segregation alters the thermodynamic driving force for vacancy attachment and detachment between $V_{n-1}$ and $V_n$, thereby establishing a direct energetic link between solute segregation and vacancy-mediated nanovoid kinetics.

Using this relation, we computed the vacancy incremental binding energies $E_b(V_{n-1}Re_m, V_1)$ for nanovoid sizes n = 1–300, with the full dataset provided in the Supplementary data. To quantify how Re coverage perturbs the stability difference between successive nanovoid sizes at fixed solute content, we introduce

$$\Delta E_B = E_B(V_n, Re_m) - E_B(V_{n-1}, Re_m), \quad (13)$$

which directly measures the Re-induced stabilization difference between adjacent nanovoid sizes. A positive $\Delta E_B$ indicates that Re segregation enhances the thermodynamic driving force for vacancy attachment, whereas a negative value implies a reduction of the vacancy trapping strength.

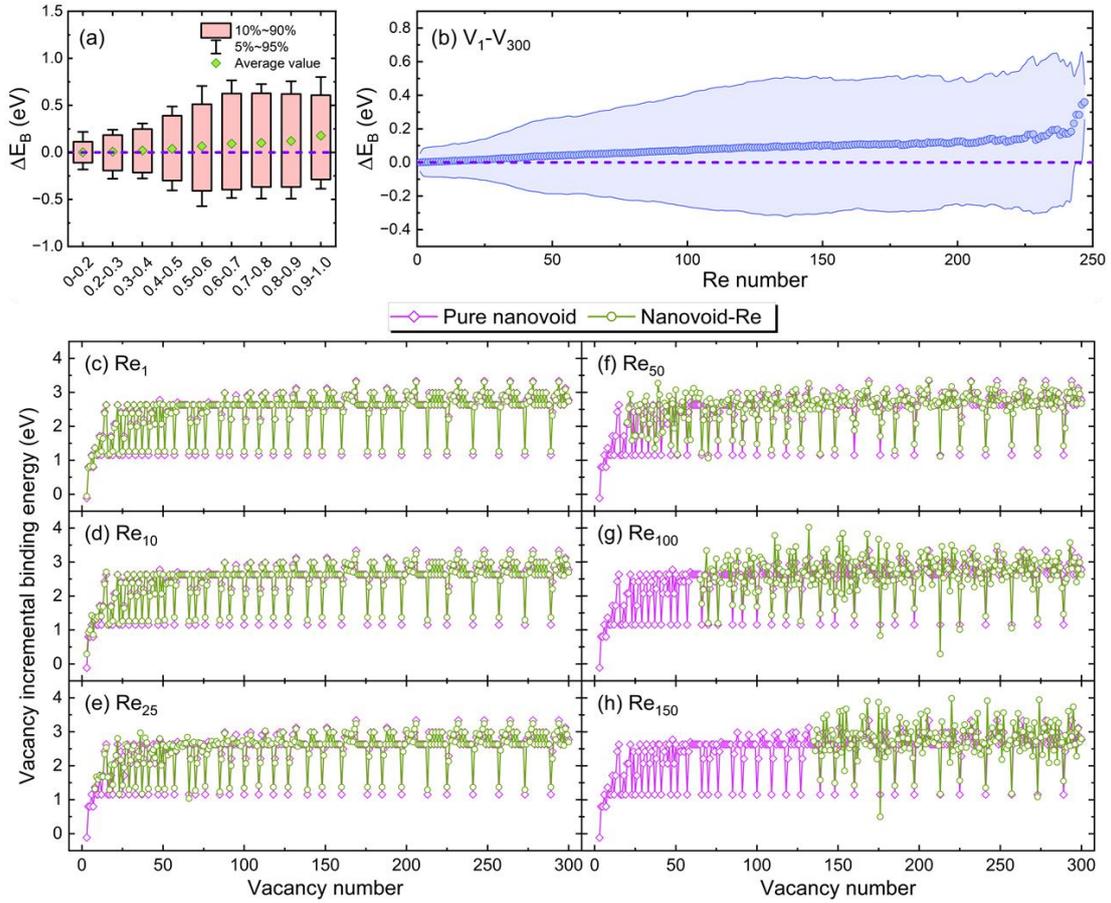

**Fig. 8.** Effect of Re coverage on the vacancy incremental binding energies of nanovoids. (a) Distribution of $\Delta E_B = E_B(V_n, Re_m) - E_B(V_{n-1}, Re_m)$ at different Re surface coverages $\theta_{Re}$. (b) Mean and standard deviation of $\Delta E_B$ as a function of Re number m. (c–h) Representative comparisons of $E_b(V_{n-1}Re_m, V_1)$ with the corresponding Re-free value $E_b(V_{n-1}, V_1)$ for m = 1, 10, 25, 50, 100, and 150, respectively, illustrating the Re-induced deviations in vacancy incremental binding energies.

Figure 8(a) summarizes the distribution of $\Delta E_B$ at different Re surface coverages $\theta_{Re}$. At low coverage ($\theta_{Re}$ =0-0.2), $\Delta E_B$ exhibits a narrow distribution, with the 5%–95% interval spanning -0.19 to 0.22 eV, indicating that sparse Re segregation introduces only a minor perturbation to the vacancy incremental binding energy of the Re-free nanovoid. As the coverage increases, however, the distribution broadens markedly, reaching –0.58 to 0.70 eV at $\theta_{Re} \in (0.5, 0.6]$, and remaining comparably wide for larger $\theta_{Re}$. The presence of both positive and negative values indicates that Re segregation can either strengthen or weaken vacancy trapping, depending on the specific nanovoids. Meanwhile, the mean value of $\Delta E_B$ increases monotonically with $\theta_{Re}$, rising from nearly zero at $\theta_{Re}$ = 0–0.2 to 0.17 eV at $\theta_{Re}$ = 0.9–1.0. This trend suggests that cases leading to enhanced vacancy binding become progressively more prevalent as the Re coverage increases. Importantly, even the most negative values of $\Delta E_B$ (about -0.6 eV) are insufficient to render the vacancy incremental binding energy negative, indicating that nanovoids remain thermodynamically attractive for vacancies even when the local effect of Re is unfavorable. Figure 8(c–h) provides representative examples at selected Re contents, where the vacancy incremental binding energies $E_b(V_{n-1}Re_m, V_1)$ are directly compared with the corresponding values for Re-free nanovoids $E_b(V_{n-1}, V_1)$, thereby visualizing how Re segregation modifies vacancy adsorption and emission energetics across different nanovoid sizes.

3.5. Comparison with existing models and empirical interatomic potentials

A range of theoretical frameworks have been developed to describe the energetics of vacancy–Re complexes in W, most notably lattice Hamiltonian approaches based on Ising or CE formalisms [44-46]. In these models, the total energy is expressed as a configurational expansion over occupation variables on a fixed parent lattice, with effective cluster interactions fitted to first-principles data and subsequently sampled using Monte Carlo techniques. Extensions to multicomponent systems and temperature-dependent thermodynamics have further expanded their applicability, while semi-empirical formulations, such as the fitted expressions proposed by Huang et al. [48], provide simplified parameterizations of incremental binding energetics for use in kinetic simulations. These relations, hereafter referred to as the Huang equations, are given by

$$E_b(V_n Re_{m-1}, Re_1) = 0.079 + 0.154\left(\frac{n}{m}\right), \quad (14)$$

$$E_b(V_{n-1}Re_m, V_1) = 0.273(\frac{m^{0.59}}{n^{0.31}}). \tag{15}$$

Despite their success, a direct comparison reveals a fundamental distinction in how the energetics of nanovoid–solute complexes are represented. Conventional CE-type approaches construct a global configurational Hamiltonian in which energetic contributions are associated with clusters defined on a parent lattice, and are therefore averaged over symmetry-equivalent embeddings. In this formalism, local environments are described implicitly through a truncated set of clusters. While, in principle, increasing the order and spatial range of clusters can systematically improve the description of local coordination effects, such an approach rapidly becomes impractical for nanovoid–solute systems due to the combinatorial growth in the number of clusters and the associated fitting complexity. This limitation is not fundamental to the CE formalism itself, but arises from its practical implementation, where the need for a tractable cluster basis restricts the explicit resolution of highly heterogeneous local environments, such as those found at nanovoid surfaces.

By contrast, the present framework is formulated in terms of physically defined incremental adsorption processes at nanovoid surfaces, in which the energetics are decomposed into nanovoid–solute and nanovoid-mediated solute–solute contributions governed explicitly by local coordination environments. As demonstrated in Section 3.1, these interactions are strongly localized within the first two neighbor shells, enabling the total energy of a nanovoid–Re complex to be reconstructed as an assembly of a finite set of distinct local-environment motifs. This formulation should therefore be viewed not as a reformulation of CE, but as an alternative representation in which the elementary energetic variables are local adsorption events rather than global occupation clusters. As a result, it provides a physically transparent description of configuration-dependent energetics while remaining computationally tractable for nanovoids of arbitrary size.

A quantitative comparison with the Huang equations (Fig. 9) further highlights these differences. For Re incremental binding energies, the Huang equations reproduce the high-coverage regime reasonably well, but systematically overestimate the binding strength at low surface coverage. For vacancy incremental binding energies, good agreement is observed only for small nanovoids, whereas a clear underestimation emerges with increasing vacancy number. These deviations can be traced to the underlying formulation: the Huang equations depend solely on the global variables n and m, and therefore neglect the configurational diversity inherent to nanovoid surfaces.

As shown in the present dataset, the incremental binding energy can vary substantially among configurations with identical (n, m), particularly at low Re coverage where spatial arrangements strongly influence nanovoid-solute and solute-solute interactions. The absence of structural descriptors in the Huang formulation thus precludes an accurate description of these variations, leading to systematic discrepancies outside its original calibration regime.

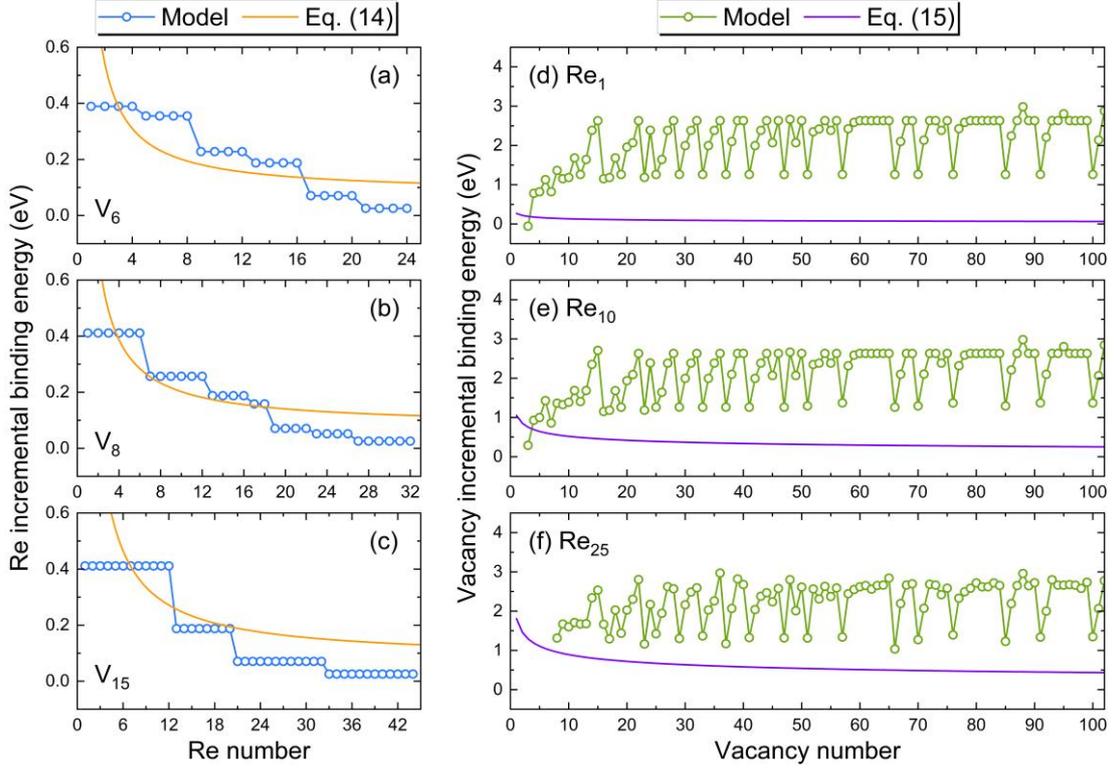

**Fig. 9.** Comparison between the Huang equations and the present model predictions for incremental binding energies in vacancy–Re complexes. (a–c) Re incremental binding energies for $V_6$, $V_8$, and $V_{15}$, respectively. (d–f) Vacancy incremental binding energies for $Re_1$, $Re_{10}$, and $Re_{25}$, respectively.

We further benchmarked several widely used empirical W–Re interatomic potentials [61-64] against the present model (Fig. 10). While all potentials qualitatively reproduce the staircase-like evolution of Re incremental binding energies, significant quantitative inconsistencies are observed. The YC1 and YC2 potentials overestimate binding energies at both low and high Re numbers while underestimating them in the intermediate regime, whereas the Setyawan and Bonny potentials underestimate the binding energies for relatively low Re numbers but overestimate them at higher Re numbers. Notably, the transition between overestimation and underestimation frequently coincides with the steps in the adsorption sequence, indicating that these potentials struggle to resolve the subtle energy differences between competing surface occupation motifs. For vacancy incremental binding energies, all potentials capture the general size-dependent trend but systematically overestimate the binding strength, particularly for small nanovoids. In several cases, anomalous peaks appear at specific void sizes, suggesting non-physical overbinding that would artificially enhance vacancy absorption and suppress emission, effectively leading to a spurious "super-trapping" behavior.

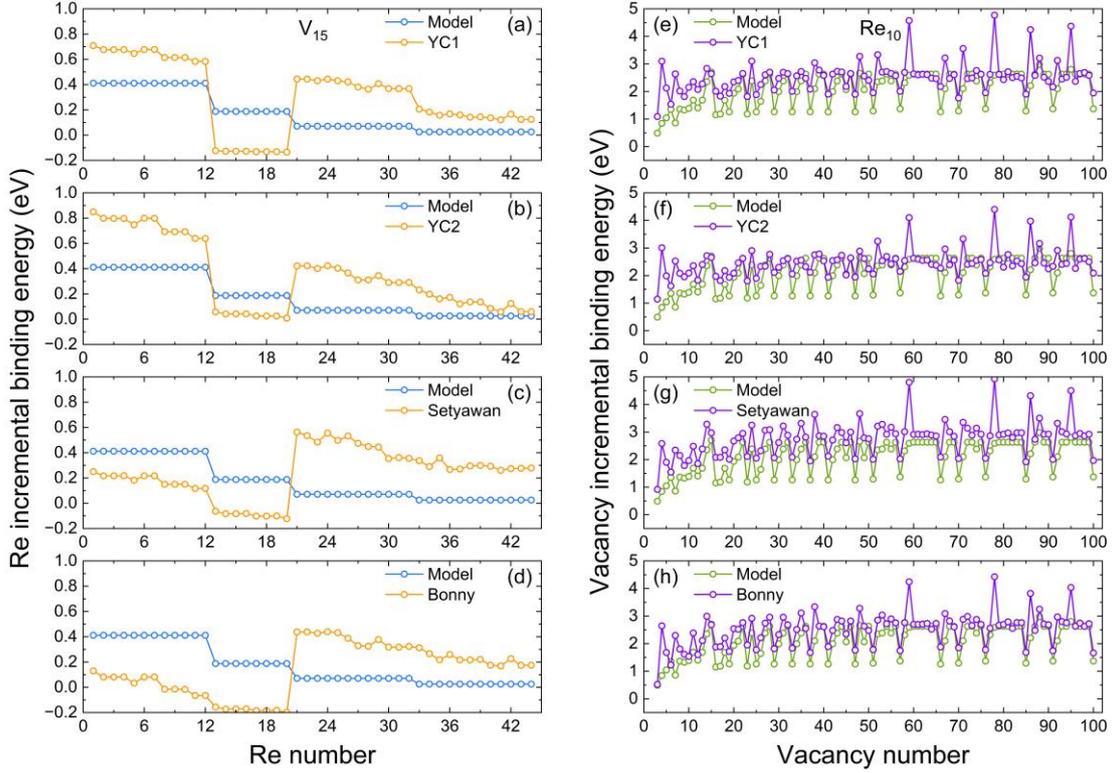

**Fig. 10.** Comparison of incremental binding energies predicted by empirical W–Re interatomic potentials and the present model. (a–d) Re incremental binding energies for $V_{15}$ calculated using the YC1 [61], YC2 [62], Setyawan [63], and Bonny [64] potentials, respectively, together with the present model predictions. (e–h) Vacancy incremental binding energies for $Re_{10}$ calculated using the YC1, YC2, Setyawan, and Bonny potentials, respectively, together with the present model predictions.

Taken together, these comparisons highlight that the key limitation of existing approaches lies not in the representation of composition alone, but in the absence of an explicit link between energetics and local coordination environments at defect surfaces. By resolving the energetics at the level of discrete adsorption events and their local environments, the present framework captures both the configurational variability and the hierarchical nature of solute segregation, thereby providing a physically transparent and quantitatively reliable description of nanovoid–Re interactions across length scales.

3.6. Model extension and comparison with experiments

Building on the local-environment framework established above, the structures and energetics of nanovoid–solute complexes can be evaluated across a wide range of sizes and compositions in a unified and computationally efficient manner. By formulating the problem in terms of incremental adsorption processes governed by short-range coordination environments, the approach effectively decouples the underlying physics from the combinatorial complexity associated with large configurational spaces. As demonstrated in Sections 3.1–3.3, the total binding energy of a nanovoid–Re complex can be reconstructed from a finite set of local-environment motifs, while the size-dependent configurational search strategy (EE, SA, and GA) enables reliable identification of stable structures across different size regimes. For nanovoids beyond the explicitly sampled regime ($V_{1-300}$), the normalized surface-coverage parameter $\theta_{Re}$ provides a coarse-grained yet accurate description of incremental binding energetics, thereby extending the predictive capability of the framework to substantially larger systems.

A key implication of this formulation is that the dominant energetics of solute segregation at nanovoid surfaces are controlled by local coordination environments within the first two neighbor shells. This pronounced locality is not specific to Re, but reflects a more general characteristic of solute adsorption at metallic void surfaces. To examine the transferability of the framework, we extended the analysis to Os and Ta, which are also relevant transmutation or alloying elements in W. As shown in Fig. 11(a–d) (see Supplementary Tables S2 and S3 for detailed data), configurations with identical 1NN and 2NN local coordination environments exhibit nearly identical binding energies across all three solute species, indicating that the local-environment features capture the essential physics of solute–defect interactions independent of chemical identity. Once calibrated using a limited set of DFT data, the same motif-based representation and machine-learning strategy can therefore be applied to different solutes without explicit enumeration of the exponentially growing configurational space. The resulting predictions for Os and Ta adsorption on $V_{50}$, $V_{150}$, and $V_{300}$ nanovoids (Fig. 11(e–j)) further demonstrate the robustness and scalability of the approach.

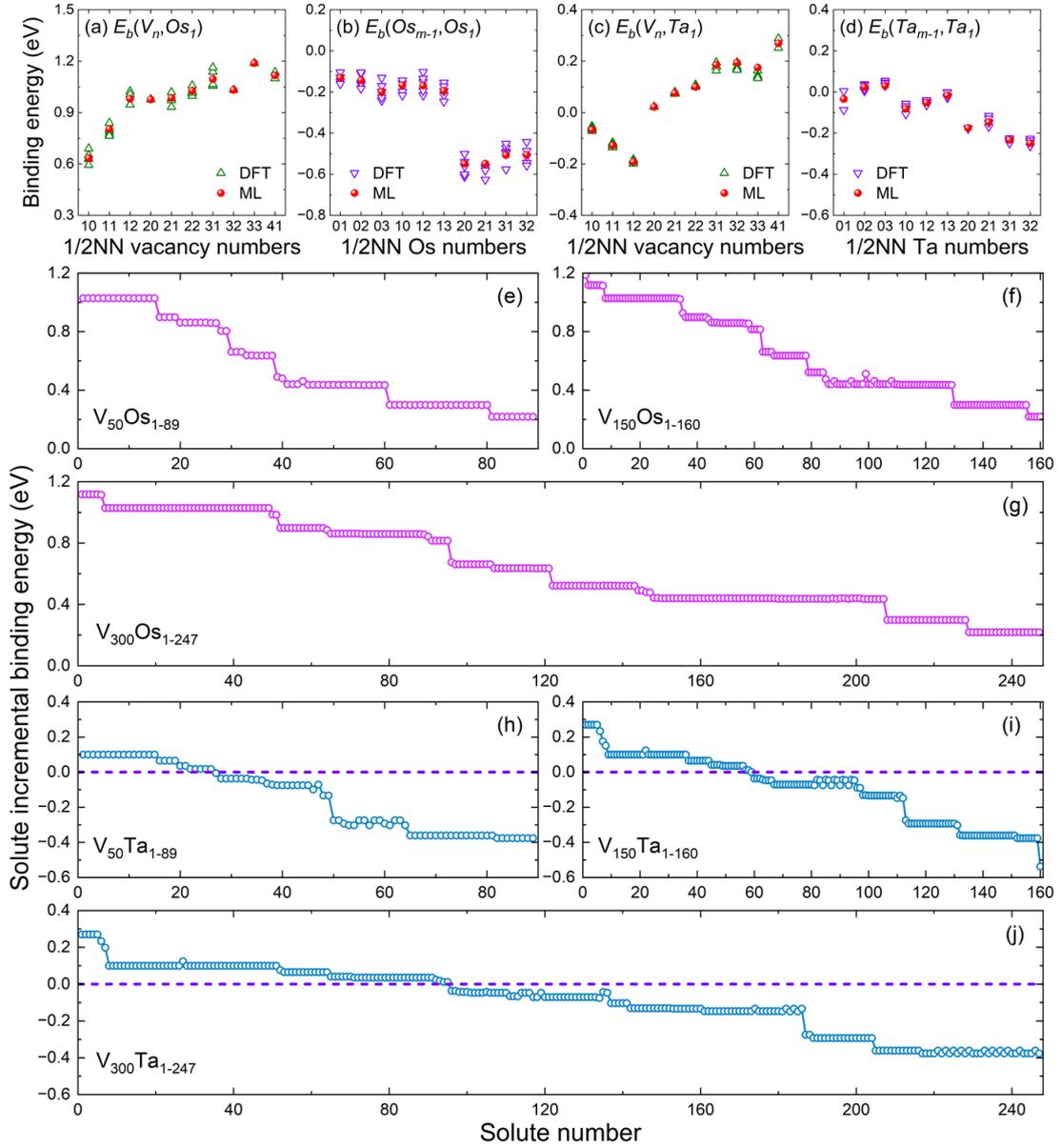

**Fig. 11.** Local-environment dependence and predicted incremental binding energies of Os and Ta on W nanovoids. (a) $E_b(V_n, Os_1)$ as a function of first- and second-nearest-neighbor (1/2NN) vacancy numbers. (b) $E_b(Os_{m-1}, Os_1)$ as a function of 1/2NN Os numbers. (c) $E_b(V_n, Ta_1)$ as a function of 1/2NN vacancy numbers. (d) $E_b(Ta_{m-1}, Ta_1)$ as a function of 1/2NN Ta numbers. (e–g) Incremental binding energies of Os sequentially adsorbed on the surfaces of $V_{50}$, $V_{150}$, and $V_{300}$ nanovoids, predicted using the GA-based model. (h–j) Incremental binding energies of Ta sequentially adsorbed on the surfaces of $V_{50}$, $V_{150}$, and $V_{300}$ nanovoids, predicted using the GA-based model.

The predicted solute segregation behavior exhibits clear chemical specificity. As shown in Figs. 4, 7(d–f), and 11(e–j), Re and Os display a strong tendency to decorate nanovoid surfaces and to subsequently form clustered configurations, whereas Ta shows a much weaker affinity for nanovoid adsorption. This contrast arises directly from the balance between nanovoid–solute attraction and solute–solute interactions encoded in the incremental binding energetics. For Re and Os, the favorable nanovoid–solute interactions dominate at low coverage and remain sufficiently strong to sustain progressive adsorption, while the evolving solute–solute interactions give rise to the characteristic hierarchical filling behavior. In contrast, for Ta, the comparatively weaker binding suppresses stable surface occupation, leading to negligible decoration. These predictions are in good agreement with available experimental observations [6, 10, 12, 13, 67]. Characterization studies of neutron-irradiated W have consistently reported the formation of Re-enriched shells around voids, often accompanied by localized Os enrichment, with segregation occurring already at relatively low damage doses and becoming more pronounced with increasing irradiation exposure. By contrast, under comparable conditions, no clear evidence for Ta decoration at nanovoid surfaces has been observed within current experimental resolution. It is worth noting, however, that some experiments report anisotropic, needle-like Re/Os-enriched features surrounding voids [5, 12]. While the present framework captures the thermodynamic driving forces for solute segregation and clustering, such anisotropic morphologies are likely influenced by irradiation-induced kinetic processes, including defect fluxes associated with vacancies and self-interstitial atoms, as well as the evolution of dislocation loops.

## 4. Conclusions

In this work, we established a physically transparent, accurate, and transferable framework for the structure–energy mapping of nanovoid–solute complexes, using nanovoid–Re complexes in W as a representative case. We showed that solute segregation at nanovoid surfaces can be rigorously decomposed into direct nanovoid–solute interactions and nanovoid-mediated solute–solute interactions. Both are governed primarily by first- and second-nearest-neighbor local coordination motifs, and identical motifs give rise to nearly identical energetics. Based on first-principles data, we trained GBDT models to establish quantitative mappings between diverse local motifs and their corresponding energetics, thereby enabling the energetics of arbitrary nanovoid–solute complexes to be reconstructed from a finite set of local motifs. We further developed a size-dependent configurational-search framework that employs EE,

SA, and GA for small, medium-sized, and large complexes, respectively, enabling the efficient identification of thermodynamically stable structures across a broad size range. Building on this framework, we generated a large structure–energy database, uncovered the staircase-like segregation behavior of Re, and derived a simple criterion based on Re surface coverage for rapid energy prediction over a wide range of nanovoid sizes. The framework also clarifies the link between Re segregation and vacancy-mediated nanovoid evolution and provides useful benchmarks for existing models and empirical interatomic potentials. Extensions to Os and Ta further support the generality of the local-motif concept, and the predicted solute segregation behavior at nanovoids is consistent with a range of experimental observations. Overall, this work provides key insights into the formation and evolution of nanovoid–solute complexes in metals, and it delivers reliable benchmarks and energetic inputs for interatomic-potential development and large-scale simulations. These advances are important for understanding and predicting defect evolution in metals.

## CRediT authorship contribution statement

**Kang-Ni He:** Writing – original draft, Writing – review & editing, Visualization, Methodology, Investigation, Formal analysis, Data curation, Conceptualization. **Xiang-Shan Kong:** Writing – review & editing, Validation, Supervision, Resources, Project administration, Methodology, Funding acquisition, Conceptualization. **Jie Hou:** Writing – review & editing, Validation, Resources. **Chang-Song Liu:** Writing – review & editing, Validation, Project administration. **Zhuo-Ming Xie:** Writing – review & editing, Validation, Supervision, Resources, Project administration, Funding acquisition.

## Declaration of competing interest

The authors declare that they have no known competing financial interests or personal relationships that could have appeared to influence the work reported in this paper.

## Acknowledgements

This work was supported by the National Natural Science Foundation of China (No.: 52571083) and the National Key R&D Program of China (Grant No. 2019YFE03110200).